\documentclass[preprintnumbers,prd,floatfix,twocolumn,amsmath,amssymb,nofootinbib]{revtex4}

\usepackage{latexsym}
\usepackage{epsfig}
\usepackage{amssymb}
\usepackage{amsmath}
\usepackage{dcolumn}
\usepackage{bm}
\usepackage{graphicx}
\usepackage{amsfonts,ulem,wrapfig}
\usepackage{multirow}
\usepackage{subfig}

\begin{document}

\title{Dynamics of cosmological scalar fields}

\author{Nicola Tamanini}
\email{n.tamanini.11@ucl.ac.uk}
\affiliation{Department of Mathematics,
University College London,
Gower Street, London, WC1E 6BT, UK}

\begin{abstract}
The background dynamical evolution of a universe filled with matter and a cosmological scalar field is analyzed employing dynamical system techniques. After the phenomenology of a canonical scalar field with exponential potential is revised, square and square root kinetic corrections to the scalar field canonical Lagrangian are considered and the resulting dynamics at cosmological distances is obtained and studied. These noncanonical cosmological models imply new interesting phenomenology including early time matter dominated solutions, cosmological scaling solutions and late time phantom dominated solutions with dynamical crossing of the phantom barrier. Stability and viability issues for these scalar fields are presented and discussed. 
\end{abstract}


\maketitle

\section{Introduction}

Since the Nobel Prize winning discovery of a current cosmological phase of accelerated expansion was made in 1998 \cite{Riess:1998cb,Perlmutter:1998np}, the theoretical models advanced to describe this phenomenon quickly multiplied in the literature. The simplest among these is the straightforward addition of a positive cosmological constant to the Einstein field equations. Although this model fits all astronomical observations, it is in tension with particle physics prediction (the cosmological constant problem \cite{Weinberg:1988cp,Martin:2012bt}) and cosmological considerations (the coincidence problem \cite{Zlatev:1998tr}).

A way to alleviate these problems consists in letting the cosmological constant to be dynamical. This implies the introduction of some cosmological field capable of reproducing the late time accelerated behavior mimiking in this way the effects of a cosmological constant. Any physical entity which at cosmic distances provide an accelerated expansion at late times is commonly called {\it dark energy}. The simplest field having these properties is a canonical scalar field with a potential. Dark energy models of this kind go under the name of {\it quintessence} and have been largely studied in the literature \cite{Copeland:2006wr,Tsujikawa:2013fta}.

Scalar fields play an important role in cosmology since they are sufficiently simple to handle and sufficiently complicated to produce non-trivial dynamics. They are not only employed to model dark energy, but also to characterize inflation \cite{Liddle:2000cg}, dark matter \cite{Magana:2012ph}, unified dark models \cite{Bertacca:2010ct} and other cosmological features. For dark matter phenomenology it is usually required a vanishing pressure and that the speed of sound of adiabatic perturbations is sufficiently small to allow the formation of clusters. Scalar field models proposed to unify dark matter and dark energy must thus have a dynamical equation of state evolving from dust-like to dark energy-like behavior, which can be achieved with a non canonical scalar field \cite{Bertacca:2010ct}.

Generalizations and modifications of the canonical scalar field Lagrangian can also lead to more complex cosmological predictions. Extended models where the scalar field Lagrangian is a general function of both the scalar field $\phi$ and its kinetic term, are known as $k$-essence theories \cite{ArmendarizPicon:2000ah}. Within this framework it is possible to obtain not only the standard dark energy evolution but also the so called {\it phantom} regime and quintessence to phantom transition, though fatal problems always arise at the level of perturbations \cite{Vikman:2004dc,Zhao:2005vj,Caldwell:2005ai}.

A phantom scalar field is identified by an equation of state (EoS) with a negative pressure bigger than the energy density. In other words, for a scalar field EoS $p_\phi=w_\phi\rho_\phi$, the phantom regime is identified by the condition $w_\phi<-1$, which seems to be slightly favoured by astronomical observations even after Planck \cite{Xia:2013dea,Novosyadlyj:2013nya}. The first and simplest model capable of achiving such condition consists in flipping the sign of the kinetic term of a canonical scalar field \cite{Caldwell:1999ew}. However in this model the scalar field EoS parameter is never greater than $-1$ creating problems at early times where dark matter with vanishing pressure must dominate. Scalar fields which can cross the phantom barrier at $w_\phi=-1$ are usually dubbed {\it quintom} models and imply either the use of extended Lagrangians, generally instable, or of two different scalar fields \cite{Caldwell:2005ai,Cai:2009zp}.

The present work is devoted to study the background dynamical evolution of different scalar field models. Dynamical system techniques are employed to fully determine the solutions of the cosmological equations. Suitable dimensionless variables are introduced following \cite{Copeland:1997et} and the phase space dynamics is analyzed using numerical methods. Canonical and noncanonical scalar field Lagrangians are presented and their cosmological implications are discussed. The complete cosmological background dynamics of two specific noncanonical scalar fields is obtained showing that interesting phenomenology, such as early time matter dominated solutions, scaling solutions, late time phantom acceleration, super-stiff and phantom transition eras, can be achieved.

The paper has the following structure. In Sec.~\ref{sec:canonical} the canonical scalar field will be largely discussed. Its features and cosmological dynamics will be presented and analyzed in depth and the notation and conventions needed for the following sections will be introduced. In Sec.~\ref{sec:noncanonical} noncanonical scalar field Lagrangians will be considered. Perturbations instabilities will be examined and the analysis will focus on models where dynamical system techniques can be sucessfully applied. Sec.~\ref{sec:square} and Sec.~\ref{sec:phantom} will then be dedicated to the study of square and square root kinetic corrections to the canonical scalar field Lagrangian. For these simple models the full dynamical features can be obtained and the background cosmological evolution can be determined for any initial condition. The analysis of these two sections will show that a rich phenomenology can be obtained with these extended scalar fields. Finally results and conclusions will be discussed in Sec.~\ref{sec:concl}


\section{The Canonical Scalar Field}
\label{sec:canonical}

In this section we review the cosmological dynamics of a canonical scalar field following the analysis first performed in \cite{Copeland:1997et}. This will serve as an introduction to the dynamical system techniques one can apply in order to completely determine the cosmological evolution of specific models. Moreover this section will be helpful in defining notation and conventions.

The action of a minimally coupled canonical scalar field is given by
\begin{align}
S = \int d^4x\sqrt{-g} \left[\frac{R}{2\kappa^2}+\mathcal{L}_\phi +\mathcal{L}_m\right] \,,
\label{014}
\end{align}
where $g$ is the determinant of the metric, $R$ is the Ricci scalar, $\kappa^2=8\pi G/c^4$, $\mathcal{L}_m$ is the matter Lagrangian, and the scalar field Lagrangian is defined as
\begin{align}
\mathcal{L}_\phi = -\frac{1}{2}\partial\phi^2-V(\phi) \,,
\label{012}
\end{align}
with $\partial\phi^2=\partial_\mu\phi\partial^\mu\phi$ and $V$ a general potential for $\phi$. The variation with respect to $g_{\mu\nu}$ produces the following gravitational equations
\begin{align}
G_{\mu\nu} = \kappa^2 \left(T_{\mu\nu}+\partial_\mu\phi\partial_\nu\phi -\frac{1}{2}g_{\mu\nu}\partial\phi^2 -g_{\mu\nu}V\right) \,,
\label{001}
\end{align}
where $G_{\mu\nu}=R_{\mu\nu}-1/2g_{\mu\nu}R$ is the Einstein tensor and $T_{\mu\nu}$ the matter energy momentum tensor. The variation with respect to $\phi$ gives the Klein-Gordon equation
\begin{align}
\Box\phi-\frac{\partial V}{\partial\phi}=0\,,
\label{002}
\end{align}
with $\Box\phi=\nabla_\mu\nabla^\mu\phi$.

In what follows we will analyze the background cosmological evolution of this model. The metric tensor will be assumed to be of the Friedmann-Robertson-Walker (FRW) type with vanishing spatial curvature
\begin{align}
g_{\mu\nu} = \mbox{diag}(-1,a(t)^2,a(t)^2,a(t)^2) \,,
\label{015}
\end{align}
with $a(t)$ the scale factor, while the scalar field is taken to be spatially homogeneous $\phi=\phi(t)$. The matter energy-momentum tensor will be of the perfect fluid form with $\rho(t)$ and $p(t)$ its energy density and pressure, respectively. A linear EoS $p=w\rho$, with $w$ the EoS parameter ranging from $0$ (dust) to $1/3$ (radiation), will be assumed.

With these assumptions, from the gravitational equations (\ref{001}) we obtain the Friedmann constraint
\begin{align}
3H^2=\kappa^2 \left(\rho +\frac{1}{2}\dot\phi^2+V\right) \,,
\label{003}
\end{align}
and the acceleration equation
\begin{align}
2 \dot H +3H^2 = -\kappa^2 \left(p+\frac{1}{2}\dot\phi^2-V\right) \,,
\label{005}
\end{align}
where $H=\dot a/a$ is the Hubble parameter and an overdot denotes differentiation with respect to the time $t$. On the other hand the scalar field equation (\ref{002}) gives
\begin{align}
\ddot\phi +3H\dot\phi + \frac{\partial V}{\partial\phi} = 0 \,.
\label{006}
\end{align}
The energy density and pressure of the canonical scalar field are given by
\begin{align}
\rho_\phi &= \frac{1}{2}\dot\phi^2+V \,,\label{052}\\
p_\phi &= \frac{1}{2}\dot\phi^2-V \,,
\end{align}
and its EoS parameter, defined as the ratio between its pressure and energy density, is
\begin{align}
w_\phi = \frac{p_\phi}{\rho_\phi} =\frac{\frac{1}{2}\dot\phi^2-V}{\frac{1}{2}\dot\phi^2+V} \,.
\label{032}
\end{align}
For $V\gg \dot\phi^2$ this approaches a cosmological constant EoS with $w_\phi=-1$, while for $V\ll \dot\phi^2$ this describe a stiff fluid with $w_\phi=1$.

At this point, following \cite{Copeland:1997et}, we introduce new dimensionless variables as 
\begin{align}
x^2 = \frac{\kappa^2 \dot\phi^2}{6H^2}\,,\quad y^2=\frac{\kappa^2V}{3H^2}\,, \quad \sigma^2=\frac{\kappa^2\rho}{3H^2} \,.
\label{020}
\end{align}
These variables are largely employed in scalar field cosmology since not only allow to rewrite Eqs.~(\ref{003})-(\ref{006}) as an autonomous system of equations, but can also be generalized in different contexts such as nonminimally coupled scalar fields \cite{Amendola:1999qq,Hrycyna:2013yia,Wei:2011yr,Xu:2012jf}, tachyons \cite{Aguirregabiria:2004xd}, Galileons \cite{Leon:2012mt}, phantom and quintom cosmology \cite{Chen:2008ft,Lazkoz:2006pa}, $k$-essence \cite{Yang:2010vv,DeSantiago:2012nk}, modified gravity \cite{Tamanini:2013ltp}, three-form cosmology \cite{Koivisto:2009fb,Boehmer:2011tp} and dark energy models coupled to dark matter \cite{Billyard:2000bh,Boehmer:2008av,Boehmer:2009tk}.

With the variables (\ref{020}) the Friedmann constraint (\ref{003}) becomes
\begin{align}
1=\Omega_m+\Omega_\phi=\sigma^2+x^2+y^2 \,,
\label{022}
\end{align}
where the relative energy densities are defined as
\begin{align}
\Omega_m = \frac{\kappa^2 \rho}{3H^2} \quad\mbox{and}\quad \Omega_\phi=\frac{\kappa^2 \rho_\phi}{3H^2} \,.
\label{055}
\end{align}
Eq.~(\ref{022}) can be used to replace $\sigma^2$ in favour of $x^2$ and $y^2$. This implies that the only dynamical variables of the system of equations will be $x$ and $y$. Also, since $\sigma^2\geq 0$ due to the assumption $\rho\geq 0$, the constraint
\begin{align}
x^2+y^2\leq 1 \,,
\end{align}
will always hold. If in addition one assumes the potential energy $V$ to be greater than zero, then $y\geq 0$ and the phase space of the variables $(x,y)$ reduces to the upper half unit disk.

At this point it is possible to convert the cosmological equations into an autonomous system of equations if one further specifies the potential $V$.
If this is exponential, for example
\begin{align}
V(\phi) = V_0 e^{-\lambda\kappa\phi} \,,
\label{004}
\end{align}
with $V_0>0$ and $\lambda$ arbitrary parameters, then the phase space will remain two dimensional. If instead one choses a power-law potential, then the phase space becomes three dimensional and the new variable
\begin{align}
z = -\frac{1}{\kappa V} \frac{\partial V}{\partial\phi} \,,
\end{align}
needs to be introduced. In this work we will only consider exponential potential of the kind (\ref{004}). With this assumption the acceleration equation (\ref{005}) and the scalar field equation (\ref{006}) lead to the two dimensional autonomous system
\begin{align}
x'&=\frac{3}{2} \left[\sqrt{\frac{2}{3}} \lambda  y^2- (w-1) x^3- x (w+1) \left( y^2-1\right)\right]\,,\label{010} \\
y'&=-\frac{3}{2} y \left[(w-1) x^2+(w+1) \left(y^2-1\right)+\sqrt{\frac{2}{3}} \lambda  x\right]\,, \label{011}
\end{align}
where a prime denotes differentiation with respect to $d\eta=Hdt$ and the variables $x$ and $y$ are functions of the dimensionless time parameter $\eta=\ln a$. Note that the dynamical system (\ref{010})-(\ref{011}) is invariant under the transformation $y\mapsto -y$, so even if we drop the $V>0$ assumption the dynamics on the negative $y$ half-plane would be a copy of the positive $y$ region. Note also that we are assuming $H>0$ in order to describe an expanding universe. However the dynamics of a contracting universe ($H<0$) would have the same features of our analysis in the negative $y$ plane switching the direction of time because of the $y\mapsto-y$ symmetry.
On the other hand the dynamical system (\ref{010})-(\ref{011}) is also invariant under the symultaneous transformation
\begin{align}
\lambda\mapsto-\lambda \quad\mbox{and}\quad x\mapsto-x \,,
\label{031}
\end{align}
which shows that opposite values of $\lambda$ lead to the same dynamics after a reflection over the $y$-axis.

The acceleration equation (\ref{005}) gives also
\begin{align}
\frac{\dot H}{H^2}=\frac{3}{2} \left[(w-1) x^2+(w+1) \left(y^2-1\right)\right]
\label{008} \,,
\end{align}
which at any fixed point $(x_*,y_*)$ of the phase space can be solved for $a$ to give
\begin{align}
a\propto (t-t_0)^{2/[3(w+1)(1-x_*^2-y_*^2)+2x_*^2]} \,,
\label{007}
\end{align}
where $t_0$ is a constant of integration.
This corresponds to a power law solution, i.e.~a solution for which the scale factor $a$ evolves as a power of the cosmological time $t$. If $x=0$ and $y=0$ the universe is matter dominated and its evolution coincides with the standard $w$-dependent scaling solution. If $x=0$ and $y=1$ the denominator of (\ref{007}) vanishes and the universe undergoes a de Sitter expansion as can be seen from (\ref{008}) which forces $H$ to be constant.
An effective EoS parameter $w_{\rm eff}$ can now be defined rewriting (\ref{007}) as\footnote{If $w_{\rm eff}<-1$ then the physical solution for the scale factor in Eq.~(\ref{007}) should be $a\propto (t_0-t)^{2/[3(1+w_{\rm eff})]}$, which implies a Big Rip at $t=t_0$.}
\begin{align}
a\propto (t-t_0)^{2/[3(1+w_{\rm eff})]} \,,
\label{009}
\end{align}
and corresponds to the EoS parameter of an effective fluid sourcing the gravitational equations, or in other words to an effective matter energy-momentum tensor. Comparing with (\ref{007}) we find
\begin{align}
w_{\rm eff} = x_*^2-y_*^2+w(1-x_*^2-y_*^2) \,.
\label{030}
\end{align}
Whenever $w_{\rm eff}<-1/3$ solution (\ref{009}) describes a universe undergoing an accelerated phase of expansion. This kind of evolution is useful to model both the inflationary early universe and the late time dark energy dominated universe. We can also have a look at how $w_\phi$ can be rewritten in terms of the variables (\ref{020}),
\begin{align}
w_\phi = \frac{x^2-y^2}{x^2+y^2} \,.
\label{056}
\end{align}
This espression tells us the equation of state of the scalar field at any given point of the phase space.

The first step one should make in order to analyze the dynamical system (\ref{010})-(\ref{011}) is to compute the critical/fixed points of the system. These are the phase space points $(x,y)$ that satisfy the conditions
\begin{align}
x'=0 \,, \qquad y'=0 \,.
\end{align}
If the system happens to be in one of these points, then there is no dynamical evolution and the universe expand according to (\ref{007}). Their existence is satisfied only if their coordinates are real and lie inside the phase space, i.e.~the upper unit half-disk in the present case. The stability conditions are computed linearizing the equations around the critical point under consideration which leads to the analysis of the eigenvalues of the Jacobian matrix
\begin{align}
\mathcal{M} =
\left(
\begin{array}{cc}
\frac{\partial f_x}{\partial x} & \frac{\partial f_x}{\partial y} \\
\frac{\partial f_y}{\partial x} & \frac{\partial f_y}{\partial y} 
\end{array}
\right) \,,
\label{050}
\end{align}
evaluated at the critical point. Here $x'=f_x(x,y)$ and $y'=f_y(x,y)$ is a compact notation for the system (\ref{010})-(\ref{011}). If the real part of both the eigenvalues is positive then the point is an unstable point, if they have different signs the point is a saddle point and if they are both negative the point is a stable point.

\begin{table*}
\caption{Critical points of the system (\ref{010})-(\ref{011}) and their properties.}
\label{tab:01}
\begin{tabular}{|c|c|c|c|c|c|c|c|}
\hline
Point & $x$ & $y$ & Existence & $w_{\rm eff}$ & Acceleration & $\Omega_\phi$ & Stability \\
\hline
\hline
\multirow{2}*{$O$} & \multirow{2}*{0} & \multirow{2}*{0} & \multirow{2}*{$\forall\;\lambda,w$} & \multirow{2}*{$w$} & \multirow{2}*{No} & \multirow{2}*{0} & \multirow{2}*{Saddle} \\ & & & & & & & \\
\hline
\multirow{2}*{$A_-$} & \multirow{2}*{-1} & \multirow{2}*{0} & \multirow{2}*{$\forall\;\lambda,w$} & \multirow{2}*{1} & \multirow{2}*{No} & \multirow{2}*{1} & Unstable if $\lambda\geq-\sqrt{6}$ \\ & & & & & & & Saddle if $\lambda<-\sqrt{6}$ \\
\hline
\multirow{2}*{$A_+$} & \multirow{2}*{1} & \multirow{2}*{0} & \multirow{2}*{$\forall\;\lambda,w$} & \multirow{2}*{1} & \multirow{2}*{No} & \multirow{2}*{1} &  Unstable if $\lambda\leq \sqrt{6}$\\ & & & & & & & Saddle if $\lambda>\sqrt{6}$ \\
\hline
\multirow{2}*{$B$} & \multirow{2}*{$\sqrt{\frac{3}{2}}\frac{1+w}{\lambda}$} & \multirow{2}*{$\sqrt{\frac{3(1-w^2)}{2\lambda^2}}$} & \multirow{2}*{$\lambda^2\geq 3(1+w)$} & \multirow{2}*{$w$} & \multirow{2}*{No} & \multirow{2}*{$\frac{3(1+w)}{\lambda^2}$} & \multirow{2}*{Stable} \\ & & & & & & & \\
\hline
\multirow{2}*{$C$} & \multirow{2}*{$\frac{\lambda}{\sqrt{6}}$} & \multirow{2}*{$\sqrt{1-\frac{\lambda^2}{6}}$} & \multirow{2}*{$\lambda^2<6$} & \multirow{2}*{$\frac{\lambda^2}{3}-1$} & \multirow{2}*{$\lambda^2<2$} & \multirow{2}*{1} & Stable if $\lambda^2<3(1+w)$ \\ & & & & & & & Saddle if $3(1+w)\leq\lambda^2<6$ \\
\hline
\end{tabular}
\end{table*}

The critical points of the system (\ref{010})-(\ref{011}) are shown in Table \ref{tab:01}. There can be up to five critical points according to the value of $\lambda$:
\begin{itemize}
\item {\it Point $O$}. The origin of the phase space corresponds to a matter dominated universe ($\Omega_m=1$) and exists for all values of $\lambda$. Of course the effective EoS matches the matter EoS, $w_{\rm eff}=w$, and thus for physically admissible values of $w$ there is no acceleration. This point is always a saddle point attracting trajectories along the $x$-axis and repelling in any other direction.
\item {\it Point $A_\pm$}. In these two points the universe is dominated by the scalar field kinetic energy ($\Omega_\phi=1$) and thus the effective EoS reduces to a stiff fluid with $w_{\rm eff}=w_\phi=1$ and no acceleration. Their existence is always garanteed and they never represent stable points. They are unstable or saddle points depending on the absolute value of $\lambda$ being greater or smaller than $\sqrt{6}$.
\item {\it Point $B$}. This point is the so called scaling solution where the effective EoS matches the matter EoS, but the scalar field energy density does not vanish. In other words we always have both $0<\Omega_\phi=3(1+w)/\lambda^2<1$ and $0<\Omega_m=1-\Omega_\phi<1$, obtaining also $w_\phi=w$. This means that the universe evolves under both the matter and scalar field influence, but it expands as if it was completely matter dominated. This solution is of great physical interest for the coincidence problem since according to it a scalar field can or could be present in the universe hiding its effects on cosmological scales. However, since we have $w_{\rm eff}=w$ there cannot be accelerated expansion. When this point exists, i.e.~for $\lambda^2\geq 3(1+w)$, it always represents a stable point attracting all the phase space trajectories.
\item {\it Point $C$}. The last point stands for the cosmological solution where the universe is completely scalar field dominated. This implies $\Omega_m=\sigma^2=0$ and $\Omega_\phi=x^2+y^2=1$ meaning that Point $C$ will always lie on the unit circle. It exists for $\lambda^2<6$ and it is a stable attractor for $\lambda^2<3(1+w)$ (i.e.~when Point $B$ does not appear) and a saddle point for $3(1+w)\leq\lambda^2<6$. The effective EoS parameter assumes the value $w_{\rm eff}=w_\phi=\lambda^2/3-1$ which implies an accelerating universe for $\lambda^2<2$. This point represents the well-known cosmological accelerated expansion driven by a sufficiently flat scalar field potential. The physical applications abound in both the early and late time universe stages. In the limit $\lambda\rightarrow 0$ this solution reduces to a de Sitter expansion dominated by a cosmological constant.
\end{itemize}

The qualitative behavior of the phase space can be divided into three regions according to the value of $\lambda^2$. In what follows we will only consider positive values for $\lambda$. The dynamics for negative values coincides with the positive one after a reflection around the $y$ axis because of (\ref{031}).

\begin{figure}
\centering
\includegraphics[width=\columnwidth]{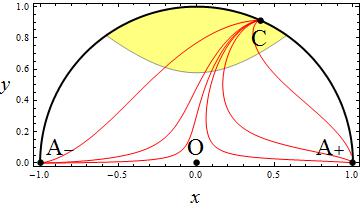}
\caption{Phase space with $\lambda=1$ and $w=0$. The global attractor is Point $C$ which represents an accelerating solution. For values $\lambda^2>2$ Point $C$ would lie outside the acceleration region (yellow/shaded) and would not be an inflationary solution.}
\label{fig:01}
\end{figure}

If $\lambda^2<3(1+w)$ there are four critical points. Points $A_\pm$ are both unstable node, while Point $O$ is a saddle point. The global attractor is Point $C$ which represents an inflationary cosmological solution if $\lambda^2<2$. The portrait of the phase space is depicted in Fig.~\ref{fig:01} where the values $\lambda=1$ and $w=0$ have been choosen. The yellow/shaded region delimits the zone of the phase space where the universe undergoes an accelerated expansion. Point $C$ always lies on the unit circle and it happens to be outside the acceleration region if $\lambda^2>2$.

\begin{figure}
\centering
\includegraphics[width=\columnwidth]{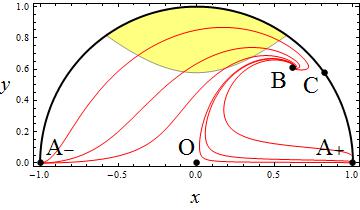}
\caption{Phase space with $\lambda=2$ and $w=0$. The global attractor is Point~$B$ where the universe expands as it was completely matter dominated, while Point~$C$ is a saddle point.}
\label{fig:02}
\end{figure}

In the range $3(1+w)\leq\lambda^2<6$ there are five critical points in the phase space. Points $A_\pm$ and $O$ still behaves as unstable nodes and saddle point respectively. The global attractor is now Point $B$ and Point $C$ becomes a saddle point. The phase space portrait is drawn in Fig.~\ref{fig:02}. Point $B$ always lies outside the acceleration region (yellow/shaded) and thus never describe an inflationary solution. However the effective EoS parameter at this point coincides with the matter EoS parameter and thus the universe experience a matter-like expansion even if it is not completely matter dominated. This is the so called scaling solution where the scalar field energy density fills part of the universe but the resulting cosmological evolution still assumes the behavior of a matter dominated expansion.

\begin{figure}
\centering
\includegraphics[width=\columnwidth]{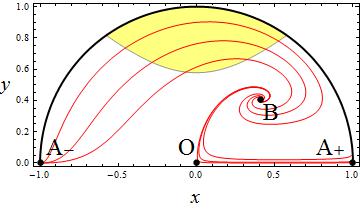}
\caption{Phase space with $\lambda=3$ and $w=0$. Point~$B$ is the global attractor describing a scaling solution with $w_{\rm eff}=w$.}
\label{fig:03}
\end{figure}

Finally if $\lambda^2\geq 6$ there are again only four critical points. Point $A_-$ is the only unstable node, while Points $A_+$ and $O$ behave as saddle points. Point $C$ does not appear anymore and the global attractor is still Point $B$, which again represents a scaling solution with $w_{\rm eff}=w$. The phase space dynamics is depicted in Fig.~\ref{fig:03}.

The cosmological dynamics of the canonical scalar field is interesting because of the appearence of late time accelerated solutions which can be employed to model dark energy and inflation. The scaling solutions are also important since allow a scalar field to hide its presence during the cosmological evolution. This situation can be used to postulate a scalar field which gives no contribution at early times but becomes relevant at late times. Unfortunately there are strong observational constraints from nucleosyntesis which force the parameter $\lambda$ to satisfy the relation $\lambda\gtrsim 9$ at early times \cite{Bean:2001wt}. Since for a late time accelerating solution a sufficiently flat potential is needed ($\lambda^2<2$), it is impossible to achieve both the scaling and accelerating regimes with a canonical scalar field and an exponential potential.

Moreover with a canonical scalar field we always have that unstable nodes of the phase space, possibly representing very early time behaviors, are associated with scalar field kinetic dominated universe. This solutions are characterized by an effective EoS approaching the stiff regime where $w_{\rm eff}=1$. Strictly speaking this value of $w_{\rm eff}$ is not physically viable at the classical level. However since these solutions appear to be relevant only at very early times in physical applications, this feature is usually ignored. As we will see in Sec.~\ref{sec:square}, with a non-canonical scalar field an early time matter dominated solution can always be obtained.


\section{The noncanonical scalar field}
\label{sec:noncanonical}

In this section we will generalize the scalar field Lagrangian $\mathcal{L}_\phi$. The canonical choice for $\mathcal{L}_\phi$ has been given in (\ref{012}) and its dynamics on cosmological scales has been investigated in full details in the previous section. In order to simplify the following equations we define
\begin{align}
X = -\frac{1}{2}\partial\phi^2 = -\frac{1}{2} g^{\mu\nu}\partial_\mu\phi\partial_\nu\phi \,.
\end{align}
The canonical choice for the scalar field Lagrangian corresponds then to $\mathcal{L}_\phi=X-V$. The most general Lagrangian containing $X$ and $\phi$ is given by $\mathcal{L}_\phi=P(X,\phi)$ where $P$ is an arbitrary function in both the variables. In cosmology such theories are known under the name of $k$-essence. They received a considerable amount of attention during the last few years because of their applications to dark energy, dark matter and inflation.

In order to reduce our analysis we will focus only on scalar field Lagrangians generally defined by
\begin{align}
\mathcal{L}_\phi= V f(B)\quad \mbox{with}\quad B=\frac{X}{V} \,,
\label{013}
\end{align}
and $f$ an arbitrary function. This includes the canonical choice if one considers $f(B)=B-1$.
In \cite{Piazza:2004df,Tsujikawa:2004dp} it has been shown that, within general relativity, the most general Lagrangian leading to cosmological scaling solutions with an exponential potential can be written as\footnote{In \cite{Piazza:2004df,Tsujikawa:2004dp} this Lagrangian was writtes as $\mathcal{L}_\phi= X\,f(B)$, however a simple redefinition of the function $f$ can bring this in the form (\ref{013}).} (\ref{013}). In addition the dimensionless variables (\ref{020}) turn out to be of great advantage if a scalar field Lagrangian is assumed as in (\ref{013}). As we will see they will permit to completely determine the cosmological dynamics of such a scalar field.
As before we will only consider the exponential potential case $V=V_0\exp(-\lambda\kappa\phi)$.

The variation of the gravitational action (\ref{014}) with the scalar field Lagrangian (\ref{013}) leads to the gravitational equations
\begin{align}
\frac{1}{\kappa^2}G_{\mu\nu} =T_{\mu\nu}+ g_{\mu\nu}\,V f+ \frac{\partial f}{\partial B} \, \partial_\mu\phi\partial_\nu\phi \,,
\end{align}
while the variation with respect to the scalar field $\phi$ gives
\begin{align}
\nabla_\mu\left( \frac{\partial f}{\partial B} \partial^\mu\phi \right) +\frac{\partial V}{\partial\phi} \left(f-\frac{\partial f}{\partial B}\,B\right) =0 \,,
\label{016}
\end{align}
Notice that these equations reduces to (\ref{001}) and (\ref{002}) for the canonical choice of the scalar field Lagrangian.

As before, the cosmological equations can be found employing the FRW metric (\ref{015}) and assuming an homogeneuos scalar field $\phi=\phi(t)$. The Friedmann constraint becomes
\begin{align}
\frac{3H^2}{\kappa^2} = \rho - V f +\frac{\partial f}{\partial B}\dot\phi^2 \,,
\label{017}
\end{align}
while the acceleration equations generalizes to
\begin{align}
2\dot H+3H^2 = -\kappa^2 \left(p+Vf\right) \,.
\label{018}
\end{align}
On the other hand the same assumptions reduce (\ref{016}) to
\begin{multline}
\left(\frac{\partial f}{\partial B}+2B\frac{\partial^2f}{\partial B^2}\right) \ddot\phi +\frac{\partial f}{\partial B}3H\dot\phi \\
-\left(f-B\frac{\partial f}{\partial B}+2B^2\frac{\partial^2 f}{\partial B^2}\right)\frac{\partial V}{\partial\phi} = 0 \,,
\label{019}
\end{multline}
where now $B=\dot\phi^2/(2V)$.

Some remarks can now be made on Eqs.~(\ref{017})-(\ref{019}). First of all we notice again that choosing $f=B-1$ reduces these equations to (\ref{003})-(\ref{006}) as expected. It is interesting to find the particular form of the function $f$ for which the contribution of the scalar field in (\ref{017}) completely disappears. This is realized for $f=\sqrt{B}$ or, in other words, for the Lagrangian $\mathcal{L}_\phi=\sqrt{XV}$. Unfortunately this particular choice also makes the first and last terms in (\ref{019}) to vanish. This implies that $\phi$ has no dynamics at all and becomes simply a constant. In any case it is worth noting that adding the $\sqrt{B}$ term to any other function $f$ does not modify the Friedmann constraint (\ref{017}) and adds a simple term $3H\sqrt{V/2}$ to the scalar field equation (\ref{019}). This features will be analyzed in more details in Sec.~\ref{sec:phantom}. Note also that it is impossible to find a scalar field Lagrangian whose contribution in the acceleration equation (\ref{018}) vanishes. This is due to the fact that $\mathcal{L}_\phi=V\,f$ and thus the vanishing of the scalar field contribution in (\ref{018}) would correspond to a zero scalar field Lagrangian.

At this point is useful to see what the Friedmann constraint (\ref{017}) looks like in terms of the variables (\ref{020}). We obtain
\begin{align}
1=\sigma^2 - y^2 f +2x^2\frac{\partial f}{\partial B} \,,
\label{021}
\end{align}
where we also have that
\begin{align}
B=\frac{x^2}{y^2} \,.
\end{align}
The Friedmann constraint (\ref{021}) determines the boundaries of the phase space described by the variables $x$ and $y$. In the canonical case (\ref{012}) this reduces to (\ref{022}) and the phase space is simply the upper-half unit circle. However if we choose a different function $f$ the phase space arising from (\ref{021}) can be considerably different from the canonical one. As a consequence we can even lose the compactness of the phase space.
In the next sections we will study what happens with different choices for the function $f$. In particular we will look for functions for which the phase space remains compact.

The expression of (\ref{018}) in terms of the $x$ and $y$ variables is given by
\begin{align}
\frac{\dot H}{H^2}= -\frac{3}{2} \left[(1+w)(1+y^2f)-2wx^2\frac{\partial f}{\partial B}\right] \,.
\label{024}
\end{align}
From this we can extract the effective EoS parameter of the universe as
\begin{align}
w_{\rm eff} = w + (w+1)y^2f-2wx^2 \frac{\partial f}{\partial B} \,.
\end{align}
We can also find the EoS of the scalar field. The energy density and pressure of $\phi$ are given by
\begin{align}
\rho_\phi&= 2X\frac{\partial f}{\partial B}-Vf \,, \label{036} \\
p_\phi&=\mathcal{L}_\phi= Vf \,.
\end{align}
Note that $\rho_\phi$ coincides with the expression appearing in the Friedmann constraint (\ref{017}) as expected. To be physically viable at the quantum level $\phi$ must satisfy $\rho_\phi>0$. The EoS parameter of the scalar field is then
\begin{align}
w_\phi = \frac{p_\phi}{\rho_\phi} = \left(2\frac{X}{V}\frac{1}{f}\frac{\partial f}{\partial B}-1\right)^{-1} \,.
\end{align}
In the canonical case $f=B-1$ this reduces to (\ref{032}), while if $f$ is constant this simply becomes $w_\phi=-1$ describing a cosmological constant.
Finally another useful quantity in scalar field cosmology is the so called speed of sound of adiabatic perturbations. This is defined as
\begin{align}
c_s^2 = \frac{\partial p_\phi}{\partial X}/\frac{\partial\rho_\phi}{\partial X} =\left(1+2\frac{X}{V}\frac{\partial^2 f}{\partial X^2}/\frac{\partial f}{\partial X}\right)^{-1} \,,
\label{034}
\end{align}
and for physically viable cosmological models we must require $c_s^2>0$. If this condition is dropped then instabilities arise at the level of perturbations of the scalar fluid. For the canonical scalar field we find $c_s^2=1$ which implies that perturbations propagates at the speed of light.


At this point, in order to completely determine the dynamics of a specific model of noncanonical scalar field, we need to choose the form of the function $f$. Ideally we would like both to find interesting phenomenology at cosmological scales and to satisfy the physical conditions $\rho_\phi>0$ and $c_s^2>0$. A possible attempt could be $f=-\exp(B)$. This choice seems indeed to yield some interesting features as one can realize looking at the EoS parameter for the scalar field which reads
\begin{align}
w_\phi = \frac{-V}{2X+V} \quad\mbox{with}\quad \mathcal{L}_\phi=-Ve^{-X/V} \,.
\label{033}
\end{align}
Whenever $X\gg V$ we have $w_\phi\simeq 0$, while if $X\ll V$ we get $w_\phi\simeq-1$. The scalar field (\ref{033}) can thus be used to characterize a dust fluid at early times and a cosmological constant at late times. Such a field could even be used to build a unified model of dark energy and dark matter, though the effects of the scalar field when $X\gg V$ would result really small since in this limit $\mathcal{L}_\phi\simeq 0$. Another drawback of this model is given by the speed of sound (\ref{034}) which results to be
\begin{align}
c_s^2=\left(1-\frac{2X}{V}\right)^{-1} \quad\mbox{with}\quad \mathcal{L}_\phi=-Ve^{-X/V} \,.
\end{align}
As we can note, as soon as $V<2X$ we obtain $c_s^2<0$ which gives rise to instabilities at the level of perturbations. Also when $V=2X$ the speed of sound diverges. We could overlook this problem for the sake of finding interesting phenomenology for the background evolution of the universe. However, as we shall see in Sec.~\ref{sec:phantom}, dropping this assumption can actually lead to a much richer cosmology if one chooses a different model.

As we can realize the exponential Lagrangian (\ref{033}) reduces to the canonical one at first order in $X/V$, so whenever this quantity is small, which usually happens at late times in cosmology, the exponential model is well approximated by the canonical scalar field. In this case the first corrections at second order would be determined by the term $-X^2/(2V)$. However, as we have seen above, the exponential Lagrangian (\ref{033}) leads to instabilities at the perturbation level. Of course there could be another form for the function $f$ which does not introduce such problems and which reduces to the canonical case when $X/V$ is small. Corrections to the canonical Lagrangian will then be given by higher order power-law kinetic terms.
Thus instead of guessing a specific form for the function $f$, we can take a starting point based on higher order (kinetic) corrections to the canonical Lagrangian. This will allow us to analyze models which both resemble the canonical scalar field at late times and are sufficiently simple to handle so that one can determine the complete cosmological dynamics of the scalar field.

We will then consider models which gives (kinetic) power-law corrections to the canonical case characterized by the function $f=B-1+\xi B^n$ with $\xi$ and $n$ two real parameters. The corresponding Lagrangian reads
\begin{align}
\mathcal{L}_\phi = X-V+\xi\,V \left(\frac{X}{V}\right)^n \,,
\label{035}
\end{align}
which is well defined only considering $n>0$.
The corrections to the canonical case are defined by the parameter $n$. For example, if $n=2$ we have that the next-to-first order corrections are of the square type, while if $n=3$ these are of the cubic type. If instead $n<1$ then we get corrections also at late times and the model does not reduce to a canonical scalar field.

The energy density (\ref{036}) and speed of sound (\ref{034}) for this model become
\begin{align}
\rho_\phi &=V+X+(2 n-1) \xi\,V \left(\frac{X}{V}\right)^n \,, \label{038}\\
c_s^2&= \frac{X+\xi  n V \left(\frac{X}{V}\right)^n}{X+\xi  n (2 n-1) V \left(\frac{X}{V}\right)^n} \,. \label{039}
\end{align}
If we assume $n> 1/2$ and $\xi\ge 0$ these quantities are always positive and finite and thus physically viable. The case $n=1/2$ is of particular interest and will be treated in Sec.~\ref{sec:phantom}, while the value $\xi=0$ yields back the canonical case.
The scalar field EoS parameter is given by
\begin{align}
w_\phi = \frac{X-V+\xi\, V \left(\frac{X}{V}\right)^n}{X+V+\xi \, (2 n-1)\,V \left(\frac{X}{V}\right)^n} \,,
\label{040}
\end{align}
and reduces to $-1$ for $V\gg X$ and to $1/(2n-1)$ for $V\ll X$ given $n>1$. This model allows for a late time cosmological constant-like EoS while the early time value of (\ref{040}) is determined by the parameter $n$. Note that for $n>1$ the scalar field EoS at early times is always positive and smaller than $1$.

In the next sections we will focus on the cases $n=2$ and $n=1/2$. The first one follows the philosophy of recovering a canonical scalar field at late times and will be studied in Sec.~\ref{sec:square}. The second one will introduce modifications at both early and late times and the phenomenology at cosmological scales will result much different and richer than the canonical one as we will see in Sec.~\ref{sec:phantom}.


\section{Square kinetic corrections}
\label{sec:square}

This section will be devoted to the dynamical analysis of background cosmologies arising from a scalar field described by Lagrangian (\ref{035}) with $n=2$:
\begin{align}
\mathcal{L}_\phi = X-V+\xi\,\frac{X^2}{V} \,.
\label{049}
\end{align}
The parameter $\xi$ will be allowed to take any real value. If $\xi=-1/2$ this model approximates the exponential model (\ref{033}) at second order in the late time small quantity $X/V$. However $\xi$ must be positive for physically viable models. In fact the energy density (\ref{038}) and sound speed (\ref{039}) reduce to
\begin{align}
\rho_\phi &= X+V+3 \xi \frac{X^2}{V} \,,\\
c_s^2 &= \frac{V+2 \xi  X}{V+6 \xi  X} \,.
\end{align}
These two quantities are always positive, for all values of $X$ and $V$, only provided $\xi>0$. Moreover we notice that the speed of sound of adiabatic perturbations reduces to one when $V\gg X$ and to $1/3$ when $V\ll X$. At early times the perturbations travels at one third of the speed of light. The EoS parameter of the scalar field (\ref{040}) becomes
\begin{align}
w_\phi = \frac{X V-V^2+\xi  X^2}{X V+V^2+3 \xi  X^2} \,.
\end{align}  
Interestingly this reduces to $-1$ when $V\gg X$ and to $1/3$ when $V\ll X$, implying that the scalar field acts as relativistic matter at early times and as an effective cosmological constant at late times. This feature signals that the model we are working with can lead to a physically sensible phenomenology at cosmological scales.

The cosmological equations (\ref{017}) and (\ref{018}) for this model are given by
\begin{align}
\frac{3 H^2}{\kappa ^2}&=\rho +V+\frac{1}{2}\dot\phi^2+\frac{3\xi}{4}  \frac{\dot\phi^4}{V} \,,\label{037}\\
3 H^2+2 \dot H&=- \kappa ^2\left(p+\frac{1}{2} \dot\phi^2- V+\frac{\xi}{4} \frac{\dot\phi^4}{V}\right) \,, \label{042}
\end{align}
while the scalar field equations (\ref{019}) becomes
\begin{align}
\left(1+3 \xi\frac{\dot\phi^2}{V}\right)\left(\ddot\phi+3 H \dot\phi\right)+\left(1-\frac{3 \xi}{4} \frac{\dot\phi^4}{V^2}\right)\frac{\partial V}{\partial\phi}=0 \,.
\label{043}
\end{align}
To determine the complete dynamics of these equations, we now employ the dimensionless variables (\ref{020}). The Friedmann constraint (\ref{037}) can then be written as
\begin{align}
\sigma^2+x^2+y^2+3 \xi\frac{x^4}{y^2}=1 \,,
\label{041}
\end{align}
where now the relative scalar field energy density is given by
\begin{align}
\Omega_\phi = x^2+y^2+3\xi\frac{x^4}{y^4} \,.
\label{054}
\end{align}
The Friedmann constraint (\ref{041}) can again be used to replace $\sigma^2$ in all the other cosmological equations. This will permit us to write the dynamical equations as an autonomous system in the variable $x$ and $y$, exactly as we did for the canonical field in Sec.~\ref{sec:canonical}. In addition, given that $\sigma^2>0$ due to the assumption $\rho>0$, the Friedmann constraint (\ref{041}) reduces the phase space to be compact and delimitated by the close geometric curve defined by $x^2+y^2+3 \xi x^4/y^2=1$. The boundary of the phase space now depends on the parameter $\xi$: the larger is $\xi$ the smaller is the phase space. The shape of the allowed phase space region can be observed in Fig.~\ref{fig:11}, where the right column shows the phase space for different values of $\xi$. The boundary of the region always presents an edge in the origin, but otherwise is smooth. The phase space becomes the upper half unit disk if $\xi\rightarrow 0$ as expected, while it reduces to the $y$-axis for $\xi\rightarrow \infty$.

With the dimensionless variables (\ref{020}) we can now write Eqs.~(\ref{042}) and (\ref{043}) as the dynamical system
\begin{align}
x'&= \frac{1}{2 y^2 \left(6 \xi  x^2+y^2\right)}\Big[18 \xi ^2 (1-3 w) x^7 \nonumber\\
&\quad -3 x y^4 \left(6 \xi x^2 (w+1)+(w-1)\left(x^2-1\right)\right) \nonumber\\
&\quad +3 \xi  x^3 y^2 \left((7-9 w) x^2+6 w-\sqrt{6} \lambda x+2\right) \nonumber \\
&\quad +y^6\left(\sqrt{6} \lambda -3 (w+1) x\right)\Big] \,,\label{044}\\
y'&= \frac{1}{2 y}\Big[3y^2 (w+1)\left(1-y^2\right)+3 \xi  (1-3 w) x^4 \nonumber\\
&\quad -xy^2 \left(\sqrt{6} \lambda +3 (w-1) x\right) \Big] \,. \label{045}
\end{align}
Note that this system is invariant under the relation $y\mapsto-y$, which implies that the dynamics on the negative $y$ half plane is symmetric to the one in the upper half plane, exaclty as it happens for the canonical scalar field. This again means that even if one drops the $V>0$ assumption, the dynamics of the whole system can be determined by just the $y>0$ analysis. Of course one always has to assume $V\neq 0$ in order for the model to not become singular. Also the transformation (\ref{031}) leaves the system (\ref{044})-(\ref{045}) unchanged, meaning that opposite values of $\lambda$ lead to the same dynamics after a reflection over the $y$-axis, again as it was in the canonical case.
Note also that given $y>0$ and $\xi>0$ the system (\ref{044})-(\ref{045}) is never singular. Moreover the origin can be taken to be part of the phase space since in the limit $x,y\rightarrow 0$ the system remain well defined as can be proved in polar coordinates\footnote{Defining $x=r\cos\theta$ and $y=r\sin\theta$ the limit $r\rightarrow 0$ always well-behaves but for the angles $\theta=0,\pi$ which however, corresponding to $y=0$, never happen to be part of the phase space.}.

From Eq.~(\ref{042}) we can also obtain
\begin{multline}
\frac{\dot H}{H^2} = \frac{3}{2 y^2} \Big[\xi  (3 w-1) x^4+(w-1) x^2 y^2 \\
+(w+1) y^2 \left(y^2-1\right)\Big] \,,
\end{multline}
from which we can obtain the effective EoS parameter at any critical point $(x_*,y_*)$ as
\begin{align}
w_{\rm eff} = w-(w-1) x_*^2-(w+1) y_*^2-\xi  (3 w-1) \frac{x_*^4}{y_*^2} \,.
\label{048}
\end{align}
In the origin this reduces to the matter EoS parameter and the scalar field has no effects on the cosmological evolution. On the other side if $x=0$ and $y=1$ this becomes $w_{\rm eff} =-1$ and the universe undergoes a de Sitter expansion.

\begin{table}
\caption{Critical points of the system (\ref{044})-(\ref{045}) and their properties. The coordinates of Points~$B$ and $C$ are given in the Appendix.}
\label{tab:03}
\begin{tabular}{|c|c|c|c|c|c|c|c|}
\hline
Point & $\, x\,$ & $y$ & Existence & $w_{\rm eff}$ & Accel. & $\Omega_\phi$ & Stability \\
\hline
\hline
\multirow{2}*{$O$} & \multirow{2}*{0} & \multirow{2}*{0} & \multirow{2}*{$\forall\;\lambda,\xi,w$} & \multirow{2}*{$w$} & \multirow{2}*{No} & \multirow{2}*{0} & \multirow{2}*{Unstable} \\ & & & & & & & \\
\hline
\multirow{2}*{$B$} & \multicolumn{2}{c|}{\multirow{2}*{App.}} & \multirow{2}*{Fig.~\ref{fig:10}} & \multirow{2}*{$w$} & \multirow{2}*{No} & \multirow{2}*{App.} & \multirow{2}*{Stable} \\ & \multicolumn{2}{c|}{} & & & & & \\
\hline
\multirow{2}*{$C$} & \multicolumn{2}{c|}{\multirow{2}*{App.}} & \multirow{2}*{$\forall\;\lambda,\xi,w$} & \multirow{2}*{App.} & \multirow{2}*{Fig.~\ref{fig:10}} & \multirow{2}*{1} & \multirow{2}*{Fig~\ref{fig:10}} \\ & \multicolumn{2}{c|}{} & & & & & \\
\hline
\end{tabular}
\end{table}

\begin{figure*}
\subfloat[][$w=0$]
{\includegraphics[width=0.9\columnwidth]{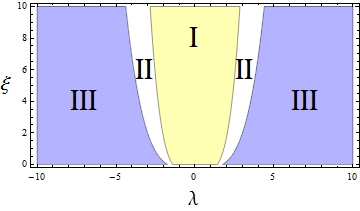}}
\qquad
\subfloat[][$w=1/3$]
{\includegraphics[width=0.9\columnwidth]{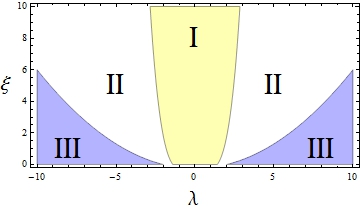}}
\caption{Existence and stability in the parameter space $(\lambda,\xi)$ of the critical points listed in Table~\ref{tab:03}. A matter EoS has been choosen in the left panel ($w=0$), while the corresponding relativistic case is shown in the right panel ($w=1/3$). Region~III denotes the existence of Point~$B$, while Point~$O$ and $C$ exist for every values of $\lambda$ and $\xi$. In region~I Point~$C$ describes an accelerating universe, while in regions~II and III it characterizes a decelerating universe. Point~$C$ is the global attractor in regions~I and II, while it is a saddle point in region~III where the global attractor is Point~$B$. The even-parity invariance of the pictures is due to the symmetry (\ref{031}) of the system (\ref{044})-(\ref{045}).}
\label{fig:10}
\end{figure*}

\begin{figure*}
\subfloat[][$\lambda=1$ and $\xi=1$]
{\includegraphics[width=0.9\columnwidth]{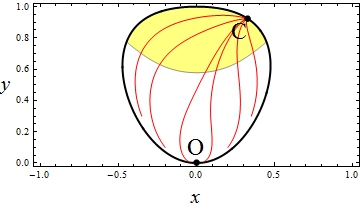}}
\qquad
\subfloat[][$\lambda=2$ and $\xi=4$]
{\includegraphics[width=0.9\columnwidth]{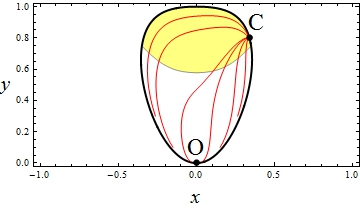}}
\\
\subfloat[][$\lambda=2$ and $\xi=1$]
{\includegraphics[width=0.9\columnwidth]{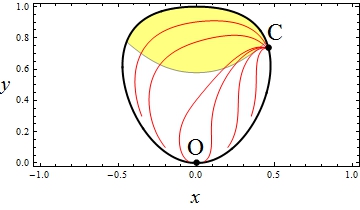}}
\qquad
\subfloat[][$\lambda=2$ and $\xi=1/2$]
{\includegraphics[width=0.9\columnwidth]{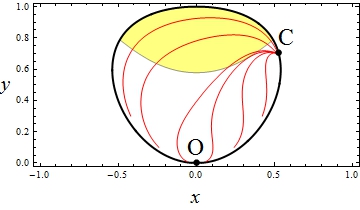}}
\\
\subfloat[][$\lambda=4$ and $\xi=1$]
{\includegraphics[width=0.9\columnwidth]{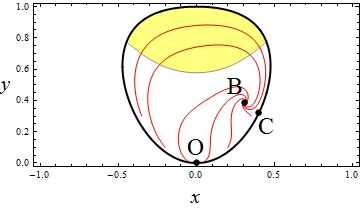}}
\qquad
\subfloat[][$\lambda=2$ and $\xi=1/10$]
{\includegraphics[width=0.9\columnwidth]{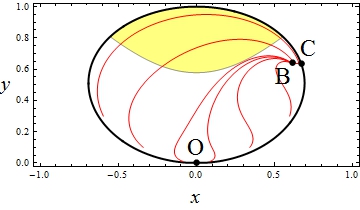}}
\caption{
Phase space of the dynamical system (\ref{044})-(\ref{045}) with the value $w=0$. The yellow region represents the zone of the phase space where the universe undergoes an accelerated expansion. In the left column the phase space is shown for different values of $\lambda$, while in the right column increasing values of $\xi$ have been displayed. Note how the boundary of the phase space changes for different values of $\xi$ while remains the same as $\lambda$ changes.
}
\label{fig:11}
\end{figure*}

The critical points for the dynamical system (\ref{044})-(\ref{045}), together with their properties, are listed in Table~\ref{tab:03}, while existence and stability are explained in Fig.~\ref{fig:10}. Assuming the origin is part of the phase space because of the considerations above, there are now only up to three critical points. Due to the high powers in $x$ and $y$ of the system (\ref{044})-(\ref{045}) the coordinate values of the critical points result quite lengthy and complicated. For this reason their explicit expression is given only in the Appendix.

\begin{itemize}
\item {\it Point $O$}. As we already noticed the origin can be taken to be part of the phase space since the system (\ref{044})-(\ref{045}) is regular at this point, as can be proved in polar coordinates. Of course it represents a matter dominated universe where the effective EoS parameter equals $w$ and $\Omega_m=1$. Interestingly the origin is now always an unstable node meaning that a completely matter dominated universe results unstable and eventually evolves to other configurations.
\item {\it Point $B$}. This point represents again a scaling solutions where $w_{\rm eff}=w_\phi=w$ but the scalar field energy density does not vanish: $0<\Omega_\phi<1$. It exists only when the parameters $\lambda$ and $\xi$ lie inside region~III of Fig.~\ref{fig:10} and it is always the global attractor of the phase space when it appears. Again since the universe evolves as it was matter dominated, this point will never characterizes an accelerating solution.
\item {\it Point $C$}. This point stands again for a completely scalar field dominated universe. In fact it always lies on the border of the phase space where $\Omega_m=\sigma=0$, $\Omega_\phi=1$ and $w_{\rm eff}=w_\phi$. However, in contrast with the canonical case, it appears in the phase space for all possible values of $\lambda$, $\xi$ and $w$. For $\lambda\rightarrow\pm\infty$ this point moves along the border of the phase space eventually approaching the origin. Regions~I and II in Fig.~\ref{fig:10} shows the values of $\lambda$ and $\xi$ where Point~$C$ is the global attractor of the phase space, while in region~III it behaves as a saddle point being Point~$B$ the global attractor. Region~I of Fig.~\ref{fig:10} represents the area in the $(\lambda,\xi)$-plane where Point~$C$ characterizes an inflationary/accelerating universe. In regions~II and III instead the effective EoS parameter in Point~$C$ is bigger than $-1/3$ and the universe undergoes a decelerating expansion.
\end{itemize}

The first feature that one notices in this model, once a comparison with the canonical case is done, is that the kinetic scalar field dominated solutions appearing as critical Points~$A_\pm$ of the system (\ref{010})-(\ref{011}), now are never part of the phase space. They are replaced by the matter dominated origin which now acts as the early time unstable solution. In this model thus, instead of having a nasty kinetic dominated solutions with $w_{\rm eff}=1$ at early time, we obtain a much more physical matter universe where $w_{\rm eff}=w$. In other words, in this model a matter dominated universe results unstable and eventually evolves to a configurations where the energy density of the scalar field does not vanish. Notice also that Point~$B$ and $C$ reduce to their correspondent canonical ones in the limit $\xi\rightarrow 0$.

We can now have a look at the complete phase space dynamics for the three different regions of Fig.~\ref{fig:10}. This has been drawn in Fig.~\ref{fig:11} where figures $(a)$ and $(b)$ represents region~I, figures $(c)$ and $(d)$ region~II and $(e)$ and $(f)$ region~III. The left column shows how the dynamics of the phase space changes as the value of $\lambda$ changes, while the right column shows how it changes as the values of $\xi$ changes. As it is clear from Fig.~\ref{fig:11}, different values of $\lambda$ do not change the shape of the phase space, while the value of $\xi$ determines the boundary, and thus the shape, of the phase space. This is of course due to the Friedmann constraint (\ref{041}) which depends on $\xi$ as we already discussed above. The yellow/shaded region represents again the zone of the phase space where the universe undergoes an accelerated expansion.

In Figs.~\ref{fig:11}~$(a)$ and $(b)$ the phase spaces for the values $\lambda=1$, $\xi=1$ and $\lambda=2$, $\xi=4$ have been depicted. The only critical point appearing beside the origin is Point~$C$, which, being inside the yellow region, characterizes an accelerating solution. All the trajectories evolves from the unstable matter dominated solution at the origin towards the scalar field dominated solution at Point~$C$ which acts as the global attractor. This dynamics well suits the phenomenology of our universe since with this parameter choice we can have a decelerated to accelerated trasition describing the dominance of dark energy over dark matter at late times and the reverse situation at early times.

The phase space dynamics for region~II of Fig.~\ref{fig:10} has been drawn in Figs.~\ref{fig:11}~$(c)$ and $(d)$ where the values $\lambda=2$, $\xi=1$ and $\lambda=2$, $\xi=1/2$ has been choosen respectively. The only two critical points in the phase space are again the origin (early time unstable solution) and Point~$C$ (late time attractor) which now lies outside the yellow region and thus describes a decelerating scalar field dominated universe. Note that, depending on initial conditions, some trajectories will still experience a stage of accelerated expansion before ending in Point~$C$. This particular evolution can thus be used to model universes with a transient inflationary era.

Finally the dynamics characterized by region~III of Fig.~\ref{fig:10} has been delineated in Figs.~\ref{fig:11}~$(e)$ and $(f)$ where the values $\lambda=4$, $\xi=1$ and $\lambda=2$, $\xi=1/2$ has been choosen respectively. There are now all three critical points in the phase space. The origin is again the early time unstable node, the global attractor is Point~$B$ representing a scaling solution and Point~$C$ is now a saddle point. Depending on initial conditions we can again have a transient acceleration era before ending at Point~$B$ with a matter-like cosmological evolution. This dynamics can be employed to build models of inflation where after the inflationary phase one obtains a graceful exit to the scaling solution.

To conclude this section we compare this model with the canonical scalar field of Sec.~\ref{sec:canonical}. Both models present cosmological scaling solutions and late time inflationary attractors. They mainly differ in the early time dynamics where instead of having kinetic scalar field dominated solutions, in the noncanonical case only the matter dominated solution appears. This feature can be used to better motivate the phenomenology of dark energy. In fact with the model presented in this section a matter dominated universe is always unstable and eventually evolves to either a scaling or a scalar field dominated solution. For the right values of the parameters $\lambda$ and $\xi$ (see Fig.~\ref{fig:10}) the late time attractor characterizes an accelerated cosmological expansion implying a dynamics describing a transition from matter to dark energy domination in accordance with the current astronomical observations.

\begin{figure}
\centering
\includegraphics[width=\columnwidth]{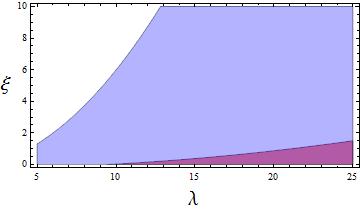}
\caption{Allowed values (purple/darker region) for early time cosmological scaling solutions of the scalar field (\ref{049}) (region~III in Fig.~\ref{fig:10}~(b)) permitted by nucleosyntesis observations. The same region appears for negative values of $\lambda$ due to symmetry (\ref{031}).}
\label{fig:12}
\end{figure}

Of course, being the model (\ref{049}) a subclass of (\ref{013}), we also obtain cosmological scaling solutions, identified with Point~$B$ in Fig.~\ref{fig:11}.
As we commented in Sec.~\ref{sec:canonical} these solutions are of great physical interest since can hide the scalar field effects on the background cosmological evolution. In the canonical case however there are strong constraints on the scalar field energy density obtained from nucleosyntesis observations which eventually impose $\lambda\gtrsim 9$. One could hope that for the scalar field (\ref{049}) the constraint on $\lambda$ would relax. Unfortunately the introduction of the square kinetic corrections, parametrize by $\xi$, does not help in this situation. As can be realized from Fig.~\ref{fig:12}, the allowed region of the $(\lambda,\xi)$-space for a viable scaling solution at early times, when $w=1/3$, is well separated from the late time acceleration region. The model (\ref{049}) thus presents the same difficulties of the canonical case for hiding a scalar field at early times which eventually becomes relevant for dark energy phenomenology at late times.

The addition of the square kinetic correction (\ref{049}) to the canonical Lagrangian complicates the resulting cosmological background equations. As a consequence the autonomous system of equations (\ref{044})-(\ref{045}) contains powerlaw terms in $x$ and $y$ up to fifth order, in contrast with (\ref{010})-(\ref{011}) where the highest order is the third. This implies that the critical points of the system are much more complex and difficult to find as it is shown in the Appendix. If one considers cubic or higher order kinetic corrections to the canonical scalar field Lagrangian the corresponding cosmological dynamical system becomes almost impossible to analyze even with numerical techniques. The model (\ref{049}) with square kinetic corrections is sufficiently simple to study and sufficiently different from the canonical case to present new phenomenology at cosmic distances, especially at early times where the scalar field kinetic dominated solutions no longer appear.
If $\xi\ll 1$ the dynamics of the scalar field (\ref{049}) approaches the corresponding canonical one. However as long as $\xi\neq 0$ the kinetic dominated solutions, corresponding to $x=\pm 1$ and $y=0$, will never appear in the phase space. Interestingly if $\xi$ is almost zero these points effectively behave as saddle points, but for a non vanishing $\xi$, no matter how small, they never characterize kinetic dominated solutions because we obtain $w_\phi=1/3$ at that points. This implies a radiation-like evolution meaning that the scalar field behaves as relativistic matter.


\section{Square root kinetic corrections}
\label{sec:phantom}

In this section we will consider square root kinetic corrections to the canonical scalar field Lagrangian. In other words we will study the model (\ref{035}) with $n=1/2$, which is simple to analyze and capable of providing interesting phenomenology at cosmological scales.
The scalar field Lagrangian of this model is then
\begin{align}
\mathcal{L}_\phi = X - V + \xi\, \sqrt{XV} \,,
\label{023}
\end{align}
where $\xi$ is again a free parameter. Recall that $V$ has been assumed to be positive and of the exponential form (\ref{004}), which means there are no inconsinstencies with the square root appearing in (\ref{023}).

Before one proceeds with the analisys of the dynamical equations, a subtlety must be taken into account. The square root in (\ref{023}) will provide terms containing $\sqrt{\dot\phi^2}$, or equivalently $\sqrt{x^2}$. This are strictly positive quantities which should be replaced by $|\dot\phi|$ or $|x|$. However this would prevent the autonomous system of equations to be differentiable in $x=0$ and the whole dynamical system analysis would be impossible since the function $f_x$ and $f_y$ of (\ref{050}) must be differentiable. For this reason we will replace $\sqrt{\dot\phi^2}$ with $\dot\phi$ in what follows, or equivalently $\sqrt{x^2}\mapsto x$. This operation can be made mathematically rigorous assuming that $\xi\mapsto s\,\xi$ with $s$ the sign of $\dot\phi$, i.e.~$s=1$ if $\dot\phi>0$ and $s=-1$ if $\dot\phi<0$. Note that we could have choose the opposite branch, i.e.~$\sqrt{\dot\phi^2}\mapsto-\dot\phi$ or $\sqrt{x^2}\mapsto-x$, however due to the symmetry (\ref{028}) of the resulting dynamical system both the choices would have result in the same cosmological dynamics.

With the scalar field (\ref{023}), the energy density (\ref{038}) and speed of sound (\ref{039}) reduce to
\begin{align}
\rho_\phi &= X+V  \,,\label{051}\\
c_s^2 &= 1 + s\frac{\xi}{2}\sqrt{\frac{V}{X}} = 1+\frac{\xi}{2}\frac{y}{x} \,. \label{053} 
\end{align}
Of course the energy density (\ref{051}) corresponds to the canonical energy density (\ref{052}).
As we already noticed in Sec.~\ref{sec:noncanonical}, we can always add a $\sqrt{B}$ term to the function $f$ on the noncanonical scalar field (\ref{013}) leaving the Friedmann constraint unmodified.
For this reason the choice (\ref{023}) yields nothing but the Friedmann constraint
\begin{align}
1=\sigma^2+x^2+y^2 \,,
\end{align}
which corresponds to the one arising in the canonical case, i.e.~(\ref{022}). This equals to say that the (gravitating) energy density of the scalar field (\ref{023}) is the same as the canonical one and that the square root term does not give any energy contribution. The scalar field relative energy density will thus be
\begin{align}
\Omega_\phi = x^2+y^2 \,,
\end{align}
which equals (\ref{055}).

The speed of sound (\ref{053}) prevents the scalar field (\ref{023}) to be physically viable. In fact it is easy to see that whenever $X=0$, or $x=0$, the speed of sound (\ref{053}) diverges giving an infinite velocity of propagation for adiabatic perturbations. Moreover if $\xi x<0$ we always obtain $c_s^2<0$ in some region of the phase space in which the scalar field will present instabilities at the perturbation level. The model (\ref{023}) results thus to be theoretically instable and non viable. However, despite all these drawbacks, in what follows we will ignore all the problems arising form Eq.~(\ref{053}). We will go on in analyzing the cosmological background dynamics implied by the scalar field (\ref{023}) showing that it is capable of producing phenomenology which cannot be obtained with the canonical scalar field and which is slightly favoured by astronomical observations.

The cosmological equations (\ref{018}) and (\ref{019}) now become
\begin{align}
2\dot H+3H^2=-\kappa^2 \left(p+\frac{1}{2}\dot\phi^2-V+\xi\,\dot\phi\,\sqrt{V}\right) \,,\label{058}\\
\ddot\phi+3H\dot\phi+3H\,\xi\,\sqrt{V}+\frac{\partial V}{\partial\phi} = 0 \,.
\end{align}
Notice that though the scalar field energy density is the same, its pressure changes due to the $\xi$ term.
This is a peculiar feature of the square root term (\ref{023}) which, despite having no gravitating energy, it yields a non zero pressure term into the acceleration equation (\ref{058}).
Moreover in the scalar field equation of motion the only modification due to $\xi$ is a new term directly coupling $H$ and the potential $V$. From these equations we obtain the following dynamical system
\begin{align}
x'&= \frac{1}{2} \Big[-3 (w-1) x^3-3 x \left[(w+1) y^2-w+1\right]\nonumber\\
&\qquad\quad +3 \sqrt{2}\, \xi\,  x^2 y+\sqrt{2}\, y \left(\sqrt{3}\, \lambda\,  y-3\, \xi \right)\Big] \,,
\label{025} \\
y'&= -\frac{1}{2} y \Big[3 (w-1) x^2+3 (w+1) \left(y^2-1\right)\nonumber\\
 &\qquad\qquad +x \left(\sqrt{6} \lambda -3 \sqrt{2} \xi  y\right)\Big] \,,
\label{026}
\end{align}
which generalizes the system (\ref{010})-(\ref{011}) with the terms containing $\xi$. Note that Eqs.~(\ref{025}) and (\ref{026}) are invariant under the symultaneous replacement
\begin{align}
\lambda\mapsto-\lambda \,,\quad \xi\mapsto-\xi \,,\quad x\mapsto-x \,,
\label{028}
\end{align}
which implies that the phase space is symmetric around the $y$-axis for opposite values of the parameters $\lambda$ and $\xi$. In the $\xi\rightarrow 0$ limit the symmetry (\ref{028}) becomes (\ref{031}). The system is also invariant under the following transformation
\begin{align}
\xi\mapsto-\xi \,,\quad y\mapsto-y \,,
\end{align}
which tells us that the dynamics in the $y<0$ half phase space equals the one in the upper half space after a redefinition of $\xi$. In the $\xi\rightarrow 0$ limit this reduces to the $y\mapsto-y$ symmetry of the canonical case.
Eq.~(\ref{024}) now reduces to
\begin{align}
\frac{\dot H}{H^2} = \frac{3}{2} \left[(w-1)\, x^2+(w+1) \left(y^2-1\right)-\sqrt{2}\, \xi\,  x y\right] \,,
\end{align}
and implies the following effective EoS parameter at any critical point $(x_*,y_*)$
\begin{align}
w_{\rm eff} = x_*^2-y_*^2+w \left(1-x_*^2-y_*^2\right)+\sqrt{2}\, \xi\,  x_* y_* \,.
\label{029}
\end{align}
Exactly as in (\ref{048}), the fact that now the parameter $\xi$ is non zero leads to new interesting phenomenology in comparison with the cosmology of the standard scalar field. The EoS parameter of the scalar field now reads
\begin{align}
w_\phi=\frac{x^2-y^2+\sqrt{2}\xi xy}{x^2+y^2} \,,
\end{align}
and differs from the canonical (\ref{056}) only by the $\xi$-term in the numerator.

\begin{table*}
\caption{Critical points of the system (\ref{025})-(\ref{026}) and their properties. The definitions of $\Omega_\phi^B$ and $Q_\pm$ are given in (\ref{057}) and (\ref{027}) respectively.}
\label{tab:02}
\begin{tabular}{|c|c|c|c|c|c|c|c|}
\hline
Point & $x$ & $y$ & Existence & $w_{\rm eff}$ & Acceleration & $\Omega_\phi$ & Stability \\
\hline
\hline
\multirow{2}*{$O$} & \multirow{2}*{0} & \multirow{2}*{0} & \multirow{2}*{$\forall\;\lambda,\xi,w$} & \multirow{2}*{$w$} & \multirow{2}*{No} & \multirow{2}*{0} & \multirow{2}*{Saddle} \\ & & & & & & & \\
\hline
\multirow{2}*{$A_-$} & \multirow{2}*{-1} & \multirow{2}*{0} & \multirow{2}*{$\forall\;\lambda,\xi,w$} & \multirow{2}*{1} & \multirow{2}*{No} & \multirow{2}*{1} & Unstable if $\lambda\geq-\sqrt{6}$ \\ & & & & & & & Saddle if $\lambda<-\sqrt{6}$ \\
\hline
\multirow{2}*{$A_+$} & \multirow{2}*{1} & \multirow{2}*{0} & \multirow{2}*{$\forall\;\lambda,\xi,w$} & \multirow{2}*{1} & \multirow{2}*{No} & \multirow{2}*{1} & Unstable if $\lambda\leq \sqrt{6}$\\ & & & & & & & Saddle if $\lambda>\sqrt{6}$ \\
\hline
\multirow{2}*{$B_\pm$} & \multirow{2}*{$\sqrt{\frac{3}{2}}\frac{(w+1)}{\lambda }$} & \multirow{2}*{$\frac{\sqrt{3}}{2 \lambda } \left(\xi \pm\sqrt{\xi ^2-2 w^2+2}\right)$} & \multirow{2}*{Fig.~\ref{fig:04}} & \multirow{2}*{$w$} & \multirow{2}*{No} & \multirow{2}*{$\Omega_\phi^B$} & \multirow{2}*{Stable} \\ & & & & & & & \\
\hline
\multirow{3}*{$C_-$} & \multirow{3}*{$\frac{\sqrt{2} \lambda -\xi  \sqrt{3 \xi ^2+6- \lambda ^2}}{\sqrt{3} (\xi ^2+2)}$} & \multirow{3}*{$\frac{ \lambda  \xi+\sqrt{6 \xi ^2+12-2\lambda ^2} }{\sqrt{3} (\xi ^2+2)}$} & \multirow{3}*{Fig.~\ref{fig:04}} & \multirow{3}*{$Q_-$} & \multirow{3}*{Fig.~\ref{fig:04}} & \multirow{3}*{1} & \multirow{3}*{Fig.~\ref{fig:04}} \\ & & & & & & & \\ & & & & & & & \\
\hline
\multirow{3}*{$C_+$} & \multirow{3}*{$\frac{\sqrt{2} \lambda+\xi  \sqrt{3\xi^2+6- \lambda ^2} }{\sqrt{3} (\xi ^2+2)}$} & \multirow{3}*{$\frac{\lambda  \xi -\sqrt{6 \xi ^2+12-2\lambda ^2}}{\sqrt{3}( \xi ^2+2)}$} & \multirow{3}*{Fig.~\ref{fig:04}} & \multirow{3}*{$Q_+$} & \multirow{3}*{No} & \multirow{3}*{1} & \multirow{3}*{Unstable} \\ & & & & & & & \\ & & & & & & & \\
\hline
\end{tabular}
\end{table*}

\begin{figure*}
\subfloat[][$w=0$]
{\includegraphics[width=0.9\columnwidth]{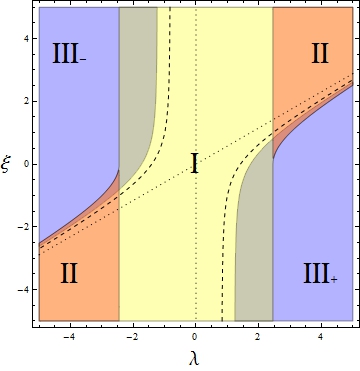}}
\qquad
\subfloat[][$w=1/3$]
{\includegraphics[width=0.9\columnwidth]{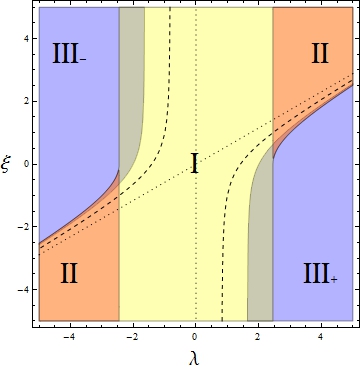}}
\caption{
Existence in the parameter space $(\lambda,\xi)$ of the critical points $B_\pm$ and $C_\pm$ listed in Table~\ref{tab:02}. A pressurless matter EoS ($w=0$) has been choosen in the left panel, while the relativistic case ($w=1/3$) is drawn in the right panel. In both cases we can identify four distinct zones. Inside zone~I the only critical point appearing in the phase space, together with the Points $A_\pm$ and $O$, is Point $C_-$, which behaves as the global attractor. Inside the disconnected zone~II both Points $C_\pm$ appear in the phase space, but Point $C_+$ is always an unstable node and Point $C_-$ again acts as the global attractor. In zone~III$_-$ we find only Points $B_-$ and in zone~III$_+$ we only have Point $B_+$. They both represent the global attractor in their respective zone. In the cross regions I/III$_\pm$ both Points $B_\pm$ and $C_-$ are present, but Points $B_\pm$ represent the global attractors, while Point $C_-$ behaves as a saddle point. The same situation happens in the crossing regions II/III$_\pm$, but now also Point $C_+$ is present and still acts as an unstable node. The connected region inside zones~I and II delimitated by the dashed lines identifies the region where Point~$C_-$ represents an accelerating solution, while in the remaining parts of zones~I and II Point~$C_-$ represents a decelerating universe. As is clear from the picture the accelerated region of Point~$C_-$ never overlaps zones~III$_\pm$, meaning that both accelerated and scaling solutions cannot exist together. Moreover the dotted lines inside the acceleration region identify the phantom regimes. The top right and bottom left parts represent region where Point~$C_-$ gives $w_{\rm eff}<-1$, while in the rest of the acceleration zone it gives $-1<w_{\rm eff}<-1/3$. Finally the odd-parity invariance of the picture is due to the transformation (\ref{028}) which leaves the dynamical system unchanged.
}
\label{fig:04}
\end{figure*}

The critical points of the system (\ref{025})-(\ref{026}) are listed in Table~\ref{tab:02}, while their existence and stability properties are expalined in Fig.~\ref{fig:04}. There are now seven possible critical points and up to six of them can appear in the phase space at the same time.

\begin{itemize}
\item {\it Point $O$}. Again the origin of the phase space formally corresponds to a matter dominated universe where $w_{\rm eff}=w$ and $\Omega_m=1$. Its properties are unmodified being always a saddle point and existing for all values of the parameters.
\item {\it Points $A_\pm$}. Also the two kinetic dominated solutions ($w_{\rm eff}=w_\phi=1$ and $\Omega_\phi=1$), labeled by Points $A_\pm$, still appear in the phase space presenting their standard behavior. In particular they are always saddle or unstable nodes depending on the absolute value of $\lambda$ being smaller or greater than $\sqrt{6}$.
\item {\it Points $B_\pm$}. These two points describe scaling solutions since in both of them $w_{\rm eff}=w_\phi=w$ and the scalar field energy density does not vanish. In fact the relative energy density of the scalar field is
\begin{align}
\Omega_\phi^B = \frac{3}{2 \lambda ^2} \left(\xi ^2\pm\xi  \sqrt{\xi ^2-2 w^2+2}+2 w+2\right) \,,
\label{057}
\end{align}
which is always between 0 and 1 when Point~$B$ exists.
Their existence is given by regions III$_\pm$ in Fig.~\ref{fig:04} and depends also on the matter EoS parameter. The smaller the value of $w$ the bigger the existence region in the $(\lambda,\xi)$ parameter space, as can be seen comparing the left and right panels of Fig.~\ref{fig:04}.
Whenever these points are present they always represent the global attractor of the phase space, but never describe accelerating solutions.
\item {\it Points $C_\pm$}. These two points represent scalar field dominated solutions and thus always lie on the unit circle being $\Omega_\phi=1$. In Fig.~\ref{fig:04} the existence of Point $C_+$ is given by the disconnected region~II, while Point $C_-$ exists in both zones~I and II. Point $C_+$ is always an unstable node, while Point $C_+$ is always the global attractor but inside the cross regions~I/III$_\pm$ and II/III$_\pm$ where it behaves as a saddle point. The effective EoS is given by $w_{\rm eff}=w_\phi =Q_\pm$, where
\begin{align}
Q_\pm = \frac{2 \lambda ^2-3( \xi ^2+2)\pm\lambda  \xi  \sqrt{6 \xi ^2+12-2 \lambda ^2}}{3 (\xi ^2+2)} \,.
\label{027}
\end{align}
This desrcibes an inflationary solution in the connected region delineated by the dashed lines as drawn in Fig.~\ref{fig:04}. Unfortunately, for positive values of $w$, this accelerating region never overlaps the existence zones of Points~$B_\pm$ meaning that inflating and scaling solutions cannot live in the same phase space.
This features appeared also in the standard case. Whenever both Points~$B$ and $C$ were present, the latter never described an inflationary solution as one can see from Fig.~\ref{fig:02}.
Finally note that, depending on the choice of parameters $\lambda$ and $\xi$, Point~$C_-$ can also describe a phantom dominated universe where $w_{\rm eff}=w_\phi<-1$. The regions in the parameter space where this happens are delimated by the two dotted line crossing the origin in Fig.~\ref{fig:04}. The top right and bottom left parts denotes phantom solutions for Point~$C_-$, while in the rest of zones~I and II we always find $w_{\rm eff}>-1$.
\end{itemize}

Note that critical Points~$B_\pm$ and $C_\pm$ reduce to Points~$B$ and $C$ of the canonical case of Sec.~\ref{sec:canonical} in the limit $\xi\rightarrow 0$.

\begin{figure}
\centering
\includegraphics[width=\columnwidth]{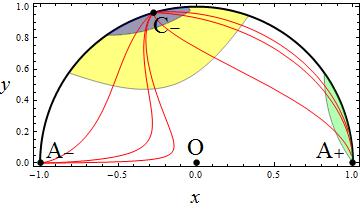}
\caption{Phase space with $w=0$, $\lambda=1$ and $\xi=1$ (region~I in Fig.~\ref{fig:04}). Point $C_-$ is the global attractor describing a phantom accelerating solution.}
\label{fig:05}
\end{figure}

\begin{figure}
\centering
\includegraphics[width=\columnwidth]{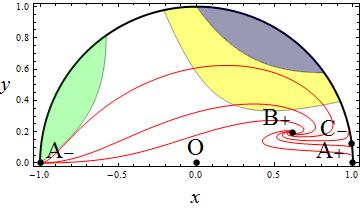}
\caption{Phase space with $w=0$, $\lambda=2$ and $\xi=-2$ (cross region~I/III$_+$ in Fig.~\ref{fig:04}). Point $C_-$ is a saddle point while Point~$B_+$ represents the global attractor describing a scaling solution with $w_{\rm eff}=w$.}
\label{fig:06}
\end{figure}

\begin{figure}
\centering
\includegraphics[width=\columnwidth]{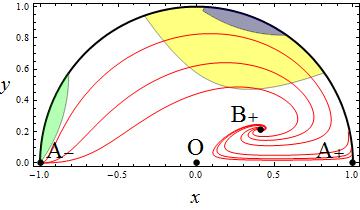}
\caption{Phase space with $w=0$, $\lambda=3$ and $\xi=-1$ (region~III$_+$ in Fig.~\ref{fig:04}). Point $B_+$ is the global attractor describing a scaling solution with $w_{\rm eff}=w$.}
\label{fig:07}
\end{figure}

\begin{figure}
\centering
\includegraphics[width=\columnwidth]{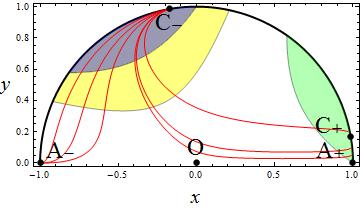}
\caption{Phase space with $w=0$, $\lambda=3$ and $\xi=2$ (region~II in Fig.~\ref{fig:04}). Point~$C_-$ is the global attractor and represents a phantom accelerated solution. Point~$C_+$ is an unstable node characterizing a super-stiff ($w_{\rm eff}>1$) solution.}
\label{fig:08}
\end{figure}

\begin{figure}
\centering
\includegraphics[width=\columnwidth]{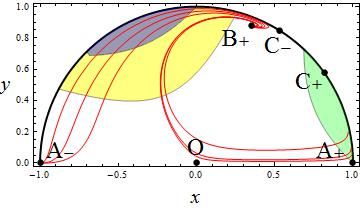}
\caption{Phase space with $w=0$, $\lambda=3.5$ and $\xi=1.5$ (cross region~II/III$_+$ in Fig.~\ref{fig:04}). Point~$B_+$ is the global attractor scaling solution, while Points~$C_\pm$ represent a saddle and unstable node respectively.}
\label{fig:09}
\end{figure}

We will now have a look at the dynamics of the phase space for values of the parameters $\lambda$ and $\xi$ representing the different zones in Fig.~\ref{fig:04}. We will restric our analysis to $w=0$ since the qualitative dynamical features do not change with other values of the matter EoS parameter. Moreover because of the symmetry (\ref{028}) we need only to study half of the parameter space, say $\lambda>0$. The remaining half will describe identical phase spaces but for a reflection $x\mapsto-x$. In Figs.~\ref{fig:05}, \ref{fig:06}, \ref{fig:07}, \ref{fig:08} and \ref{fig:09} the yellow/shaded region identifies the part of the phase space where the EoS parameter of stationary points is smaller than $-1/3$, implying an accelerated cosmological solution. The blue/dark part inside the yellow/shaded region delimitates the zone where the universe undergoes a phantom acceleration, i.e.~where $w_{\rm eff}<-1$. Finally the green/shaded region denotes the area of the phase space where the effective equations of state takes super-stiff values, i.e.~where $w_{\rm eff}>1$.

We start considering zone~I. If we choose $\lambda=1$ and $\xi=1$ the phase space looks like the one drawn in Fig.~\ref{fig:05}. Points~$A_\pm$ are unstable nodes while the origin $O$ represents the matter dominated saddle point. The global attractor is Point~$C_-$ which happens to be inside the accelerated region and thus describes an inflationary solution with $w_{\rm eff}=-11/9$. Being also inside the phantom region this value is clearly smaller than $-1$. Moreover since it lies on the unit circle it characterizes a universe completely dominated by the scalar field. If we had choosen the parameters $\lambda$ and $\xi$ to be outside the connected region delimitaded by the dashed lines in Fig.~4 but still inside zone~I, then Point~$C_-$ would have been outside the accelerated region, though still on the circle. In that case it would have described a decelerating universe dominated by the scalar field. On the other hand if we had choosen parameters inside the acceleration region, but outside the phantom region then we would have obtained an acceleration with $w_{\rm eff}>-1$. Note how the accelerated region is now different from the one in the standard case, Figs.~\ref{fig:01}, \ref{fig:02} and \ref{fig:03}. This is due to the difference between the two EoS parameters (\ref{030}) and (\ref{029}). Moreover because in Eq.~(\ref{029}) there is a dependence on $\xi$, the acceleration region will change whenever $\xi$ is different, as in the next examples.

The second zone we analyze in Fig.~\ref{fig:04} is the superposition region between zone~I and zone~III$_+$. Choosing the values $\lambda=2$ and $\xi=-2$ the phase space can be depicted as in Fig.~\ref{fig:06}. Points~$A_\pm$ and $O$ are again unstable nodes and a saddle point respectively. Point~$C_-$ is now a saddle point and the global attractor is Point~$B_+$ describing a cosmological scaling solution with effective EoS parameter matching the matter one. Point~$B_+$ clearly lies outside the accelerated region which happens to be modified with respect to the one in Fig.~\ref{fig:05}, as we discussed above.

The phase space characterized by zone~III$_+$ of Fig.~\ref{fig:04} is depicted in Fig.~\ref{fig:07} where the values $\lambda=3$ and $\xi=-1$ have been choosen. Point~$B_+$ is again the global attractor representing a cosmological scaling solution. Point~$A_-$ is still an unstable node, while Point~$A_+$ is now a saddle point exactly as the origin $O$.

In Fig.~\ref{fig:08} the portrait of the phase space for zone~II has been drawn. Now both Points~$C_\pm$ appear, one describing an unstable node ($C_+$) and the other one representing the global attractor ($C_-$) which can lie inside the phantom ($w_{\rm eff}<-1$), accelerated ($-1<w_{\rm eff}<-1/3$) or decelerated ($w_{\rm eff}>-1/3$) regions depending on the values of $\lambda$ and $\xi$. Points~$A_\pm$ are unstable nodes and Point~$O$ is a saddle point. 

Finally the phase space of the region where zones~II and III$_+$ superpose has been depicted in Fig.~\ref{fig:09}. Now Points~$C_\pm$ appear together with Point~$B_+$ representing a scaling solution. The situation is now similar to the one in Fig.~\ref{fig:06} (crossing zone~I/III$_+$): Point~$B_+$ is always the global attractor while Point~$C_-$ is a saddle point. The only difference is now Point~$C_+$ which acts as an unstable node. Points~$C_\pm$, representing the scalar field dominated solutions, always lie outside the accelerated region and thus never describes an inflationary solution. However, as it is clear from Fig.~\ref{fig:09}, before ending in Point~$B_+$ several trajectories pass through the accelerated (phantom) region, meaning that the universe undergoes a stage of accelerated (phantom) expansion before scaling as a matter dominated solution.

It is now interesting to compare the results we obtain from the model (\ref{023}) with the ones following from the canonical scalar field of Sec.~\ref{sec:canonical}. The square root term in (\ref{023}) leads to a much richer phenomenology at cosmic distances which includes phantom late time solutions, scaling solutions, new early time unstable solutions, super-stiff behavior and dynamical crossing of the phantom barrier at $w_{\rm eff}=-1$. Within this model one can not only achieve a matter to phantom transition at late times, but also phantom and super-stiff transient eras. This can be easily seen from Figs.~\ref{fig:05} to \ref{fig:09} where, depending on initial conditions, some trajectories of the phase space will cross the blue and green regions representing phantom and super-stiff behavior respectively. Thus the scalar field (\ref{023}) can describe a universe which is phantom dominated at late times instead of being only dark energy dominated as it happens in the canonical case. Despite the problems at the level of cosmological perturbations arising from Eq.~(\ref{053}), the scalar field model (\ref{023}) is actually better in agreement with the latest astronomical observations which favour a value $w_\phi<-1$ at present, though the minus one value still lies inside the 2-sigma confidence limit \cite{Xia:2013dea,Novosyadlyj:2013nya}. The scalar field (\ref{023}) can thus characterize a quintom scenario where the crossing of the phantom barrier happens at late times with the universe being nowadays dark energy dominated ($w_{\rm eff}>-1$) but evolving through a final phantom era ($w_{\rm eff}<-1$). However in order to render this a viable model of our universe one must first solve the problems arising at the level of cosmological perturbations.

\begin{figure}
\centering
\includegraphics[width=\columnwidth]{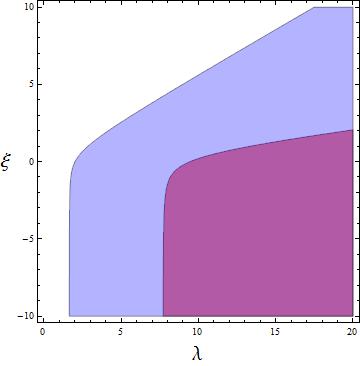}
\caption{Allowed values (purple region) for early time cosmological scaling solutions of the scalar field (\ref{023}) (region~III$_+$ in Fig.~\ref{fig:04}~(b)) permitted by nucleosyntesis observations.}
\label{fig:13}
\end{figure}

Finally, being the scalar field (\ref{023}) a subclass of (\ref{013}), we obtain again cosmological scaling solutions (Points~$B_\pm$) which, as we said, can be of great phenomenological interest. The scalar field can in fact hiding its presence at early times letting the cosmological evolution scaling as a matter dominated universe. For this to happen however we need to satisfy the constraints derived from nucleosystesis observations \cite{Bean:2001wt}. In Fig.~\ref{fig:13} the region allowed by these constraints for Point~$B_+$ in the $(\lambda,\xi)$-space is shown. For Point~$B_-$ the same region appears at opposite values of $\lambda$ and $\xi$ due to the symmetry (\ref{028}).  The introduction of the square root term in (\ref{023}) does not help in relaxing the $\lambda\gtrsim 9$ constraint of the canonical case. In fact, as can be realized from Fig.~\ref{fig:13}, the allowed region is well separated from the acceleration region of Point~$C_-$, meaning that scaling and late time accelerated solution cannot appear in the same phase space. The same happens in both the canonical case and the model of Sec.~\ref{sec:square} where the allowed region results to be much more constrained as can be understood comparing Fig.~\ref{fig:13} and Fig.~\ref{fig:12}. 



\section{Conclusion}
\label{sec:concl}

In the present work the cosmological background evolution characterized by different scalar field models has been studied. The use of dynamical system techniques has allowed to completely determine the cosmological features of canonical and noncanonical scalar fields.
After the canonical model has been reviewed in Sec.~\ref{sec:canonical}, extended scalar field Lagrangians have been presented and discussed in Sec.~\ref{sec:noncanonical}. The analysis has then focused to models whose dynamics can be completely parametrized by the dimensionless variable (\ref{020}) and which always leads to scaling solutions.

In Sec.~\ref{sec:square} a scalar field with square kinetic corrections to the canonical Lagrangian has been examined. Its late time evolution qualitatively corresponds to the canonical situations with scaling and scalar field dominated solutions, while the early time features result modified. In particular the scalar field kinetic dominated solutions no longer appear in the phase space of this model. The early time behavior is now characterized by a matter dominated solution, which is better in agreement with a radiation/dark matter dominated epoch as required by observations. The model can thus be used to describe a universe where dark energy becomes important only at late times while dark matter dominates at early times. It also happens to be safe at the level of perturbations once the condition $\xi>0$ is assumed. Furthermore the phase space boundaries of the model presented in Sec.~\ref{sec:square} differ from the canonical ones. The phase space ceases to be the upper half unit disk in the $(x,y)$-plane and, remaining compact, assumes a form depending on the parameter $\xi$ as can be seen from Fig.~\ref{fig:11}. This is an interesting mathematical feature which implies that the variables (\ref{020}) can lead to different phase space boundaries depending on the scalar field Lagrangian one chooses.

Sec.~\ref{sec:phantom} has been devoted to the study of the cosmological consequences of a scalar field models with square root kinetic corrections to the canonical Lagrangian. The background dynamics of this model presents a richer phenomenology with respect to the canonical case. The early time behavior results similar to the canonical one, though super-stiff ($w_{\rm eff}>1$) transient regions always appear in the phase space. What changes more is the late time evolution where phantom dominated solutions, dynamical crossings of the phantom barrier and new scaling solutions emerge in the phase space. This model can thus be used to describe a late time dark energy dominated universe capable of dynamically crossing the phantom barrier ($w_{\rm eff}=-1$) as the astronomical observations slightly favour. Moreover we can achieve transient periods of super-acceleration ($\dot H>0$) where the universe expands only for a finite amount of time. These solutions are characterized by the trajectories that cross the phantom region in Figs.~\ref{fig:05} to \ref{fig:09} and can be employed to build phantom models of inflation. The drawbacks arise of course at the level of pertubations where instabilities of the scalar field always appear. Until these problems are unsolved the scalar field model of Sec.~\ref{sec:phantom} cannot be seriously employed to describe physical universes.


\acknowledgments

The author would like to thank Christian B\"ohmer, Marco Bruni, Emmanuel Saridakis and David Wands for useful discussions and comments on the paper.

\appendix

\section{}

In this appendix we will provide the coordinate values of the critical points of the system (\ref{044})-(\ref{045}). For the sake of simplicity we will assume  $w=0$ in what follows.

Point~$B$ is identified by the coordinates
\begin{align}
x_B&=\frac{1}{\lambda}\sqrt{\frac{3}{2}}\,, \\
y_B&=\frac{\sqrt{3}}{4|\lambda|}\left(1+\sqrt{4 \xi +1}\right)^{1/2} \,,
\end{align}
while Point~$C$ assumes the complicate values
\begin{align}
x_C&=\frac{\Delta ^{2/3}+4 \lambda \Delta^{1/3} -36 \xi  \left(\lambda ^2+4\right)+7 \lambda ^2-36}{3 \sqrt{6}(4 \xi +1) \Delta^{1/3} }\,,\label{046} \\
y_C&=\frac{1}{\sqrt{6}}\Bigg[3 (x_C^2+1)-\sqrt{6} \lambda  x_C \nonumber\\
 &\qquad\quad+\sqrt{36 \xi  x_C^4+\left(3 x_C^2-\sqrt{6} \lambda  x_C+3\right)^2}\Bigg]^{1/2} \,,\label{047}
\end{align}
where
\begin{multline}
\Delta=54 \left(48 \xi ^2+8 \xi -1\right) \lambda +(10-216 \xi )\lambda ^3 \\
+9 (4 \xi +1) \Big[36 \xi  \left(\lambda ^6-12 \lambda ^4+24 \lambda ^2+64\right) \\
+5184 \xi ^2 \lambda ^2-\left(\lambda ^2-6\right)^2 \left(3 \lambda^2-16\right)\Big]^{1/2}
 \,.
\end{multline}
Note the complexity of the coordinates of Point~$C$ where the best expression one can obtain for $y_C$ is only in terms of $x_C$.
Finally to obtain the effective EoS parameter and $\Omega_\phi$ for Point~$C$ one should insert expressions (\ref{046})-(\ref{047}) into (\ref{048}) and (\ref{054}) respectively. These values has not been displayed due to their lenght.

\end{document}